\documentclass[a4paper,11pt]{article}
\pdfoutput=1 % if your are submitting a pdflatex (i.e. if you have
\usepackage{jinstpub} % for details on the use of the package, please
\usepackage{lineno}

\title{\boldmath A new geometry of scintillating crystals with Strip SiPMs: a PET detector with precise position and time determination}

\author[a,1]{K.\,Doroud,\note{Corresponding author.}}
\author[a]{Z.\,Liu,}
\author[a,b,c]{and M.C.S.\,Williams} 

\affiliation[a]{CERN Geneva, Switzerland}
\affiliation[b]{INFN, Bologna, Italy}
\affiliation[c]{Gangneung-Wonju National University, Gangneung, South Korea}

\emailAdd{katayoun.doroud@cern.ch}

\abstract{Measurement of the Time-of-Flight (TOF) of the 511\,keV gammas brings an important reduction of statistical noise in the PET image, with higher precision time measurements producing clearer images.  Scintillating crystals are used to convert the 511\,keV annihilation photon to an electron of $\sim$\,511\,KeV energy via the photoelectric effect; it is necessary to determine with precision the position and time of this conversion within the scintillating crystal. We propose using an array of crystals cut into a specific geometry discussed below; these crystals are read out by an array of strip SiPMs. This technique allows individual time measurements of the first arriving photo-electrons and to extract the best time resolution using a specific algorithm. The final result is a precise determination of the 3D position (that includes the depth of interaction) of the photoelectric interaction and an improved time measurement.}


\begin{document}
\maketitle
\flushbottom
%\linenumbers

%\begin{linenumbers}
\section{Introduction} 

Positron Emission Tomography (PET) is a 3D image corresponding to a distribution of radio-tracer molecule injected into the subject being scanned. This radio tracer emits a positron that annihilates with a nearby electron to form two back-to-back annihilation photons of 511\,keV.  The positron from the decay of $^{18}$F has a typical range of 1-2 mm before annihilation with an electron; thus the PET image is in the millimetre scale.  The PET image is built up by detecting these back-to-back 511\,keV annihilation photons, constructing a line that connects the detection positions of the two 511\,keV gammas; these lines are known as the line of response (LOR).  If the time of the interaction of the gamma in the detector is precisely measured, the location of the positron annihilation along the LOR is constrained; this constraint is tightened by improving the time resolution. The quality of this time measurement is expressed by the CoincidenceTime Resolution (CRT), which is defined as the Full Width Half Maximum (FWHM) of the time difference distribution of the two gamma photons; if the CTR is 10\,ps then the line segment is 1.5\,mm in length and the measurement of each pair of back-to-back annihilation photons would result in a 3-D point (rather than a line segment). However a CTR of 10\,ps implies that the time of the interaction of each annihilation photon  is measured with a time resolution of 3\,ps (sigma); such precision is currently not possible.  The goal of our work is to reach a CTR of 100\,ps; this restricts the LOR to a line segment of 1.5 cm in length.

Currently, scintillating crystals, together with Silicon Photo-Multiplier (SiPM) photosensors are one of the best choices for a PET system design, if high sensitivity and precise timing measurement are the major goals. A SiPM consists of a matrix of silicon avalanche diodes. Each diode is biased such that it operates in Geiger mode; thus a single photon can trigger a Geiger breakdown in a diode and produce a detectable signal.  For this reason the pixels are known as single photon avalanche diodes, giving the name of SPAD to these pixels; typically SPADs have dimensions in the range 10\,$\times$\, 10\,$\mu$m$^2$ to 100\,$\times$\, 100\,$\mu$m$^2$. This feature makes this device especially interesting when it is coupled to a scintillating crystal. The passage of a charged particle in a scintillating crystal produces a burst of light with a characteristic rise and fall time. The time of arrival of photons at the photo-detector is a Poisson process where the variance is equal to the mean; thus the best time resolution is obtained using the time measurement of the earliest photoelectron.  However the rise-time, the decay-time, the transport of the light within the crystal and the Gaussian time response of the SiPM itself modifies this.  In these situations, simulation models have shown that the best time resolution is given by the time of arrival of a subsequent photoelectron\,\cite{Schaart} when all effects of time jitter are taken into account. In reality, the lowest possible threshold settings (usually below the 1 photo-electron level) are used to give the best time resolution \cite{sipm-threshold}.

For a TOF-PET detector, it is of course necessary to have the best time resolution, but equally important is to have high detection sensitivity for the 511\,keV annihilation photons.  The photoelectric effect converts the 511\,keV annihilation photon to a 511\,keV electron; it is this electron that stimulates the scintillating crystal to generate a light pulse. The choice of selecting a suitable scintillator for TOF-PET depends on many factors, such as: light output, the duration of the light pulse, high Z to enhance the photoelectric effect ratio compared to Compton scattering and a short radiation length for the 511\,KeV annihilation photons.  The family of Lutetium Silicates is one of the popular choices of scintillating crystals. LYSO has a radiation length is 11.2\,mm\cite{lyso-lfs}; thus for a reasonable sensitivity, it is important to have 15\,mm (or longer) crystal length. 

\begin{figure}[!b]
\centering
\includegraphics[width=10cm]{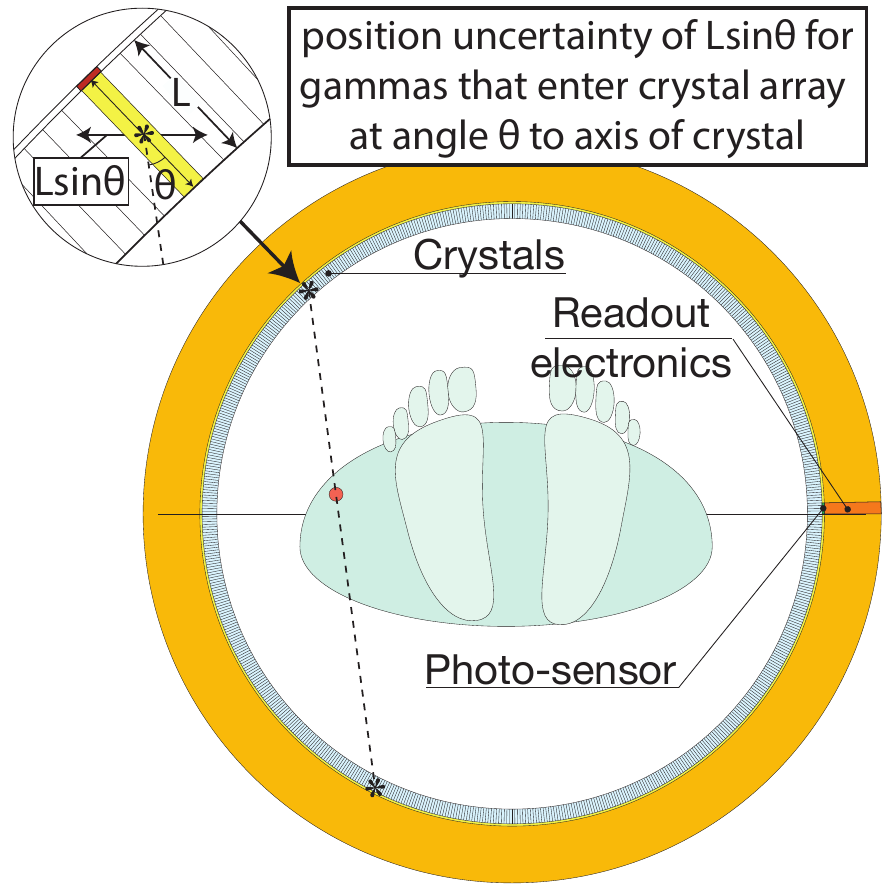}
% where an .eps filename suffix will be assumed under latex, 
% and a .pdf suffix will be assumed for pdflatex; or what has been declared
% via \DeclareGraphicsExtensions.
\caption{Typical schematic of a PET scanner consisting of crystals arranged to point to the centre. These crystals are readout by a photosensor at the far end. When an annihilation photon interacts in a particular crystal, the exact position along the  crystal axis is unknown; this introduces an uncertainty Lsin$\theta$ in the position determination (L is the length of the crystal, $\theta$ is the angle between the direction of the annihilation gamma and the axis of the crystal). This uncertainty smears the image. }
\label{parallax}
\end{figure}

As an example, reference \cite{endotof} describes a detector for a TOF-PET module with a conventional geometry; the LYSO crystals have a size of 3.5\,$\times$\,3.5\,$\times$\,15\,mm$^3$ with each crystal read out by a Hamamatsu MPPC of 3\,$\times$\,3\,mm$^2$.  The average CTR for the 4096 measured crystal pairs was 239\,ps. The best commercial TOF PET scanner is the Siemens Biograph Vision scanner \cite{SiemensTOF} with a CTR of 214\,ps.  This TOF-PET scanner is built as a ring with a diameter of 78\,cm; it uses LSO crystals of a similar geometry as in reference \cite{endotof}. It should be noted that with this geometry the position of the annihilation photon interaction along the crystal axis is not known; this limits the CTR and also will smear the final PET image since many annihilation photons will enter the detector at an angle to the crystal axis.This effect is known as the parallax error.   This is illustrated in fig.\,\ref{parallax}. If a  CTR of 100\,ps is to be realised, it is crucial to locate the position of the photoelectric interaction along the axial direction of the crystal (i.e.\,the depth of interaction). 

Previously, we have proposed a new strip geometry for the SiPM \cite{Doroud} (Hamamatsu has fabricated the Strip SiPM used by us, thus hereafter, we refer to them as Strip MPPCs (multi pixel photon counter)). The strip MPPC is read out at each end, with each end coupled to an individual TDC (time to digital converter). The time difference is related to the position of the firing SPAD along the length of the strip MPPC, while the average of the two times gives the time of the hit. The strip geometry has the advantage that the ends of the strip MPPC are at the edge of the photodetector device; this allows both the anode and the cathode to be accessed and thus a differential signal can be sent to the front-end electronics. The principles of the new strip MPPC geometry have been discussed previously\cite{Doroud}; we measured a single photon time resolution of $\sim$\,50\,ps sigma. In this paper, we discuss a method of coupling a block of scintillating crystals to this new geometric design of MPPCs. We present our preliminary and encouraging results, highlighting the high spatial resolution (i.e.\,the ability to determine the location of the annihilation photon interaction within the crystals) and the time resolution.

\begin{figure}[!b]
\centering
\includegraphics[width=11 cm]{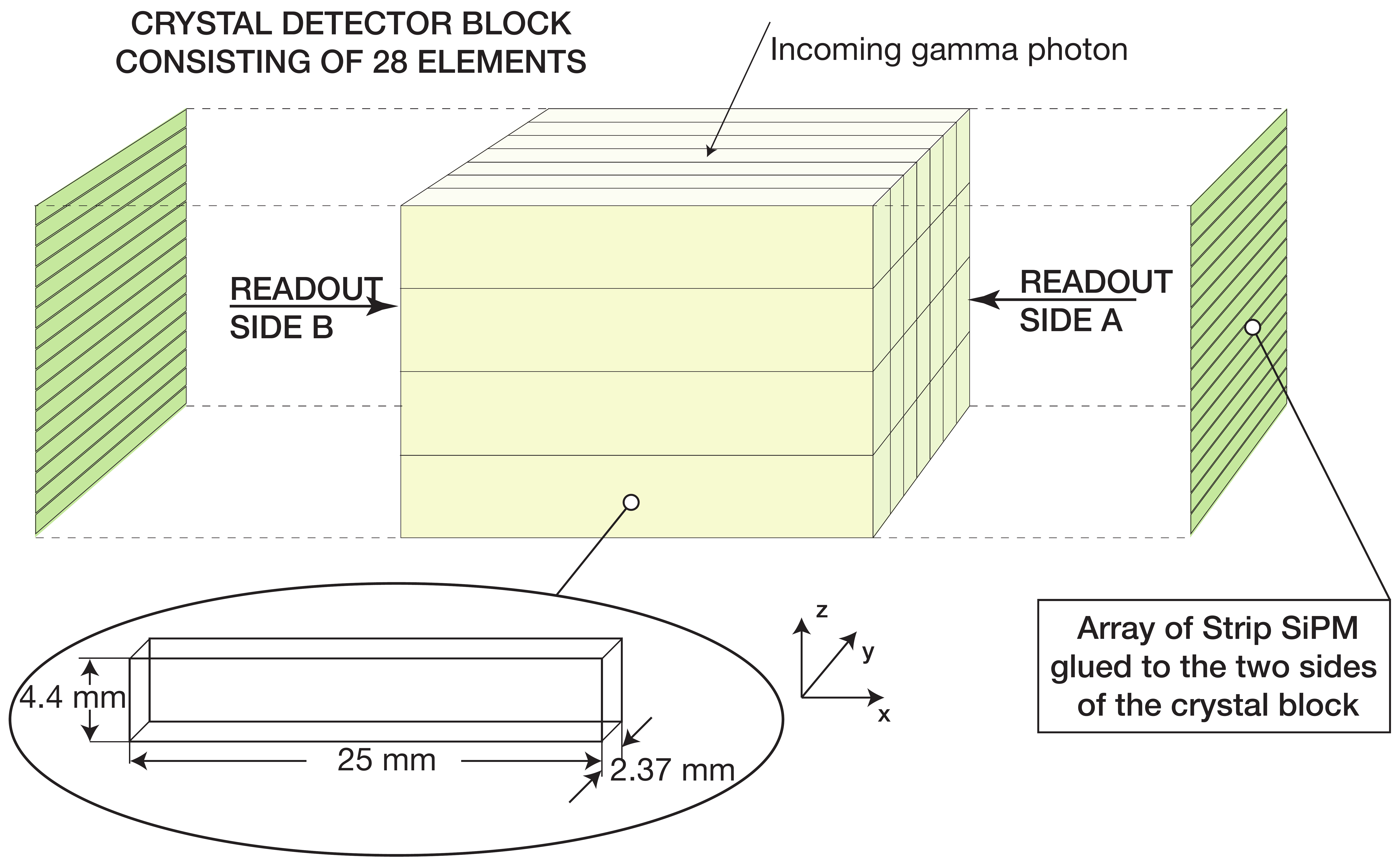}

\caption{A crystal block consists of 28 individual crystals.}
\label{schematic-slabs}
\end{figure}

\begin{figure}[!t]
\centering
\includegraphics[width=14cm]{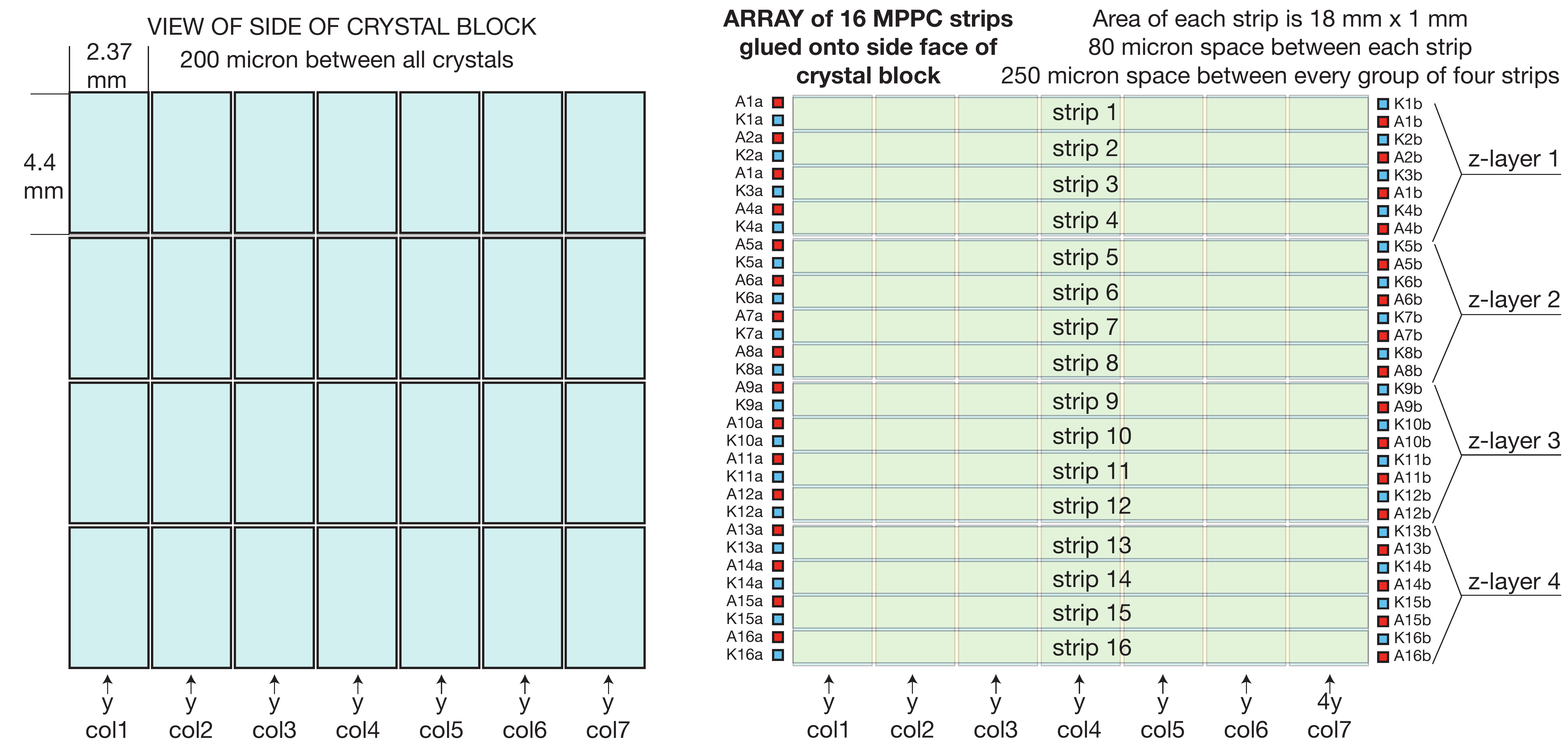}
% where an .eps filename suffix will be assumed under latex, 
% and a .pdf suffix will be assumed for pdflatex; or what has been declared
% via \DeclareGraphicsExtensions.
\caption{The view of the readout at one side of the crystal block; 16 strip MPPCs are mounted onto a substrate.  The strip MPPC have an active area of 18\,$\times$\,1\,mm$^2$ with 80\,$\mu$m between each strip MPPCs and 250\,$\mu$m between groups of four strip MPPCs. The array of strip MPPCs shown on the right hand side are glued onto crystal face.}
\label{strip-layout}
\end{figure}

\section{Detector Design}

We propose a new geometry based on the Strip MPPC \cite{patent}. Fig.\,\ref{schematic-slabs} shows a stack of these crystals assembled into a detector block. As discussed above it is necessary to measure the position of the annihilation photon interaction (including the depth of interaction) in order to utilise annihilation photons that enter at wide angles. The depth of the crystal block shown in fig.\,\ref{schematic-slabs} is 17.6\,mm, but it is divided into 4 layers (each layer has a depth of 4.4\,mm); determining the layer that contains the interaction is straight forward, thus a good depth of interaction can be determined. The dimension of each strip MPPC is 18\,$\times$1\,mm$^2$; thus every crystal element is read out by four strip MPPCs mounted on each side; this will allow individual timing of the initial burst of photons. Additionally if there is a dark count in the strip MPPC just prior to the real event, this can be recognised and rejected from the analysis.  The 200\,$\mu$m space between each crystal is filled with powdered Barium Sulphate (BaSO$_4$).

Each crystal has a dimension of 25\,$\times$\,2.37\,$\times$\,4.4\,mm$^3$.   Two Strip MPPC arrays (shown in fig.~\ref{strip-layout}) that consist of 16 strip MPPCs are attached to each side (readout side A and B) of the crystal block detector; thus every crystal element is read out by four strip MPPCs mounted on each side (i.e. a total of eight strip MPPCs are coupled to each scintillating crystal).   Within this crystal block, the position of the annihilation photon interaction is defined in the x direction by measuring the time difference and light amplitude difference between the strip MPPCs mounted on side A and B.  In the y direction, it is defined by the y-width of the crystal (this is 2.4 mm in this case) where the time difference between the two ends of the strip MPPC selects which crystal contains the interaction.  The z position (commonly referred to as the ``depth of interaction") is determined by selecting the group of 4 strip MPPCs that have a signal.  Thus with this geometry we obtain a 3D position determination with an appropriate position resolution for PET detectors; in addition, there are eight strip MPPCs with signals; these data can be used to give an enhanced time resolution.

\begin{figure}[!t]
\centering
\includegraphics[width=10cm]{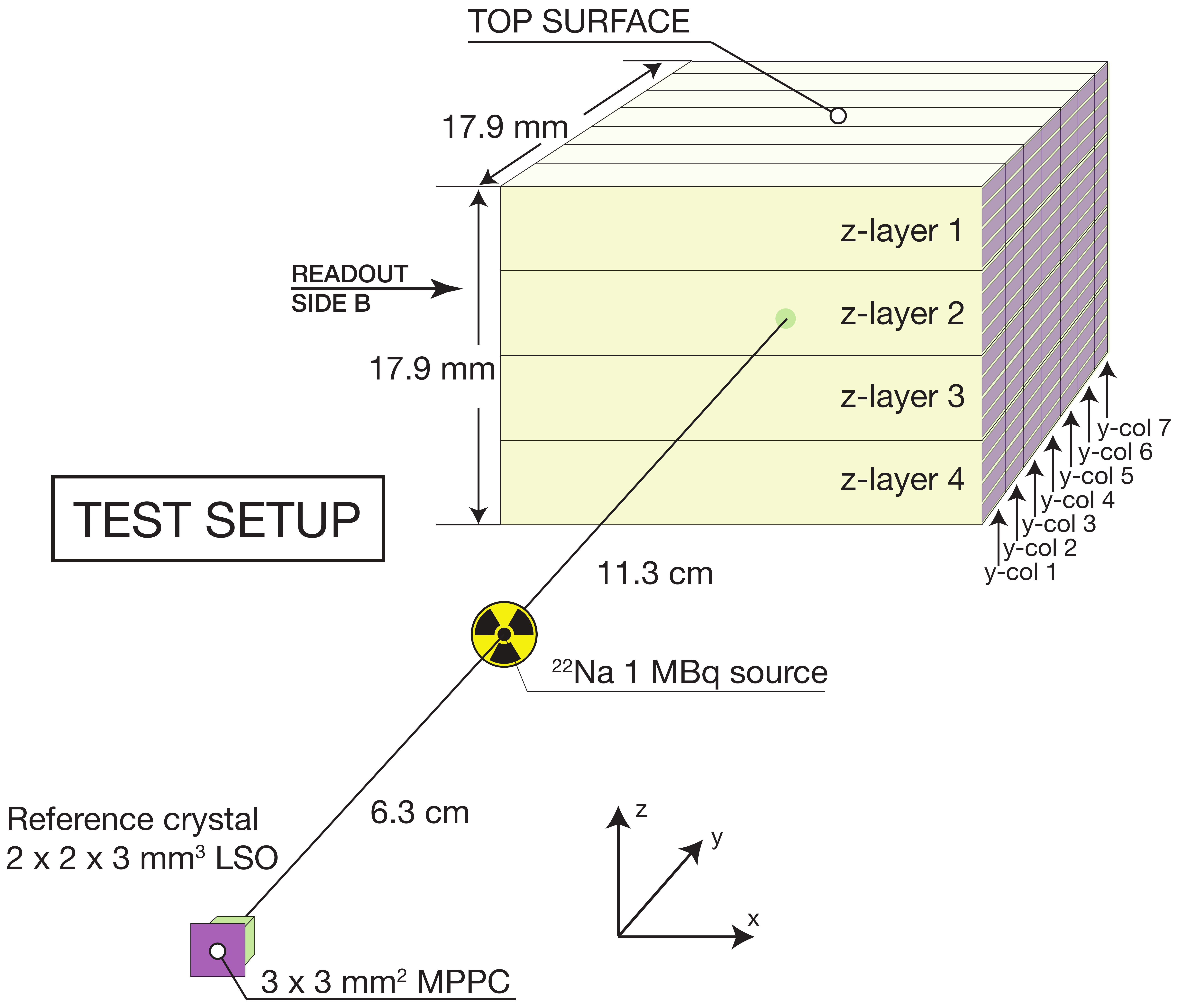}
% where an .eps filename suffix will be assumed under latex, 
% and a .pdf suffix will be assumed for pdflatex; or what has been declared
% via \DeclareGraphicsExtensions.
\caption{Test setup: on one side of a $^{22}$Na source a crystal array was mounted on a computer controlled support that could be scanned in the x-z plane; on the other side of the source, a reference crystal (2\,$\times$\,2\,$\times$\,3\,mm$^3$ LSO manufactured by Agile) is mounted ; this is read out by a 3\,$\times$\,3\,mm$^2$ MPPC. For this test, the crystal block was orientated so that the annihilation photons entered into the front surface.  When used as a TOF-PET detector, the annihilation photons will enter through the top surface.}
\label{setup}
\end{figure}

To build a TOF-PET scanner the top surface of the scintillator block shown in fig.\,\ref{schematic-slabs} would be orientated towards the subject being scanned; thus the 511\,keV annihilation photons would enter through the top surface. However for the tests presented here the annihilation photon enters the block of crystals through the front face as shown in fig.\,\ref{setup}. 

The strip MPPCs were fabricated without a protection layer of epoxy to minimise the distance between the crystal and MPPC.  The crystals were glued to the MPPCs with EP601-LV epoxy glue manufactured by Epotek. 

Two crystal arrays were tested, one fabricated from LYSO  and one fabricated with LFS.  The plots shown in this paper will be labelled LFS and LYSO where appropriate.

\section{Results from tests}

\subsection{Position and pulse height measurements}

\begin{figure}[!t]
\centering
\includegraphics[width=11cm]{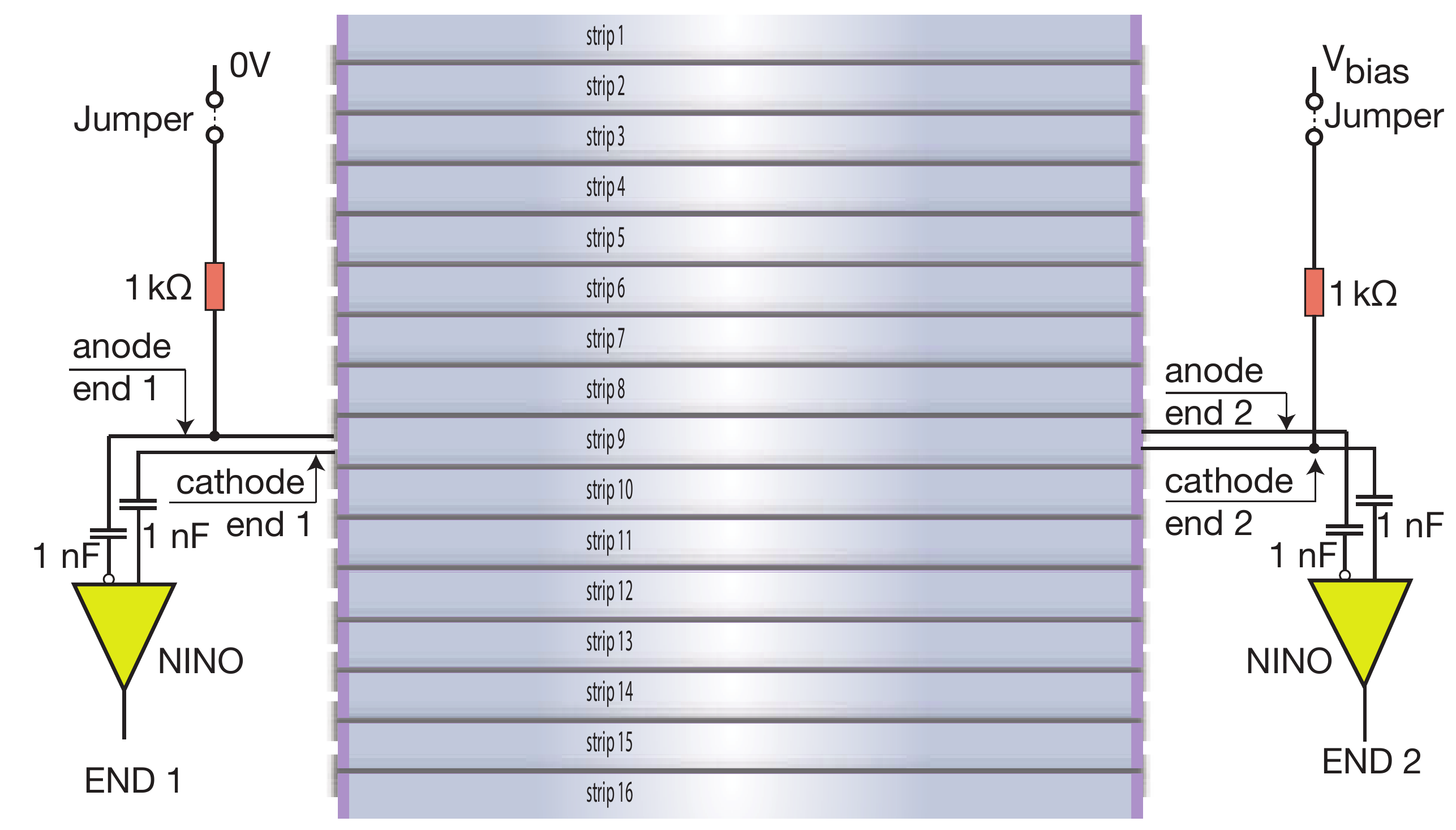}
% where an .eps filename suffix will be assumed under latex, 
% and a .pdf suffix will be assumed for pdflatex; or what has been declared
% via \DeclareGraphicsExtensions.
\caption{Circuit diagram of readout of the strip MPPC.  For clarity the circuit for just one strip MPPC is shown; however all 16 strip MPPCs were connected to readout electronics. The voltage supply and the ground connection for each strip MPPC could be disabled by removing a jumper; however all strip MPPCs were powered on for these tests.}
\label{strip-circuit}
\end{figure}

To test the performance of the crystal block and the strip MPPCs, we mounted the crystal block on one side of a $^{22}$Na source and a reference crystal (2\,$\times$\,2\,$\times$\,3\,mm$^3$ LSO (manufactured by Agile)) on the other side as shown in fig.\,\ref{setup}. This reference crystal is readout with a 3\,$\times$\,3\,mm$^2$ MPPC (type S13360-3050VE) from Hamamatsu. From previous measurements we know that the time jitter of the reference is 35\,ps sigma for 511 keV photoelectric events.  The crystal block was mounted on a computer controlled support system that could move the block in the x-z plane (see fig.\,\ref{setup}).
Each end of a strip MPPC was connected to a channel of the NINO \cite{nino} asic mounted close to the strip MPPCs.  The output pulse from the NINO front-end electronics is LVDS, with the leading edge corresponding to the time when the input signal crosses a simple threshold (usually this threshold corresponded to $\sim$\,0.3 photoelectron)\cite{sipm-threshold}. The width of this output pulse corresponds to the time-over-threshold and is thus, related to the input charge. The circuit diagram for a single strip MPPC is shown in fig.\,\ref{strip-circuit}; all 16 strip MPPCs were attached to similar power and readout circuits. Due to a shortage of readout electronics, we could only read out 16 strip MPPCs with TDCs at one time (i.e. eight strip MPPCs on each side of the crystal block); however all 32 strip SiPMs were powered for all tests.  The TDC measurements were made with a Wavecatcher \cite{wavecatcher} system.  All strip MPPCs were operated at 61\,V (6\,V overvoltage) for all tests; the reference MPPC was operated at 54.5\,V (2\,V overvoltage).

\begin{figure}[!b]
\centering
\includegraphics[width=8cm]{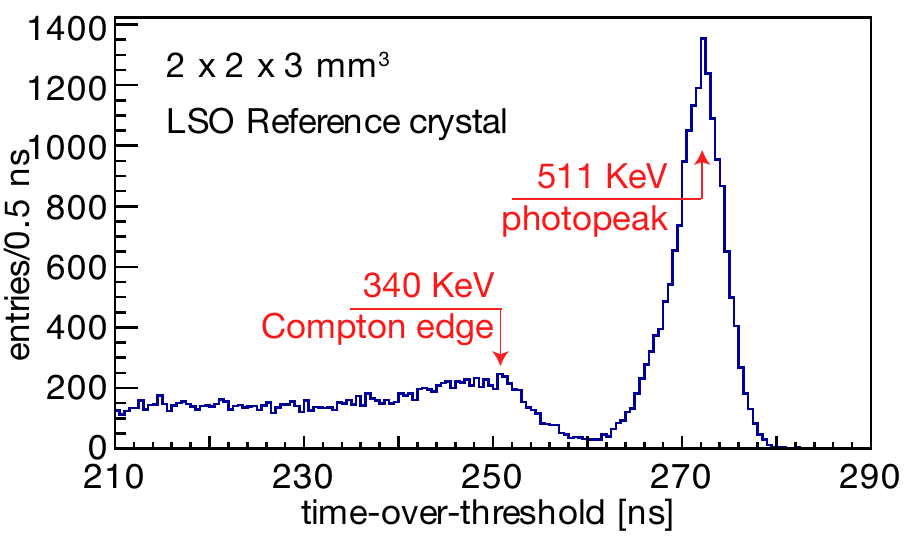}
% where an .eps filename suffix will be assumed under latex, 
% and a .pdf suffix will be assumed for pdflatex; or what has been declared
% via \DeclareGraphicsExtensions.
\caption{Time-over-threshold measured by the NINO ASIC for the LSO reference scintillator. Even though the width of the output signal from the NINO ASIC is advertised to be time-over-threshold, in reality the internal circuitry is more complex and this is not a true time-over-threshold.  However we can get an estimate of the energy scale by recognising the 340\,KeV Compton edge and 511\,KeV photopeak. }
\label{charge-ref}
\end{figure}

 \begin{figure}[!b]
\centering
\includegraphics[width=15cm]{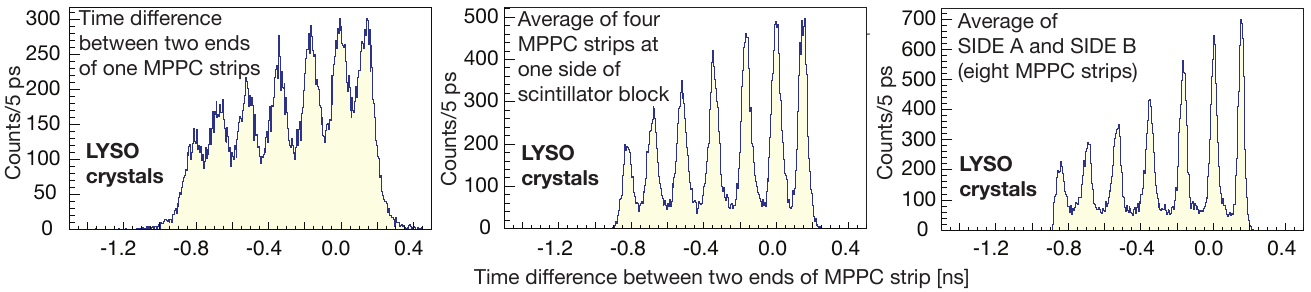}
% where an .eps filename suffix will be assumed under latex, 
% and a .pdf suffix will be assumed for pdflatex; or what has been declared
% via \DeclareGraphicsExtensions.
\caption{The time difference between the two ends of the strip MPPCs for a single MPPC, for four strips (corresponding to one z-layer) on one side of the scintillator block,  and the average of all eight MPPC strips for the two sides of the scintillator block.}
\label{time-diff-1}
\end{figure}

\begin{figure}[!t]
\centering
\includegraphics[width=10cm]{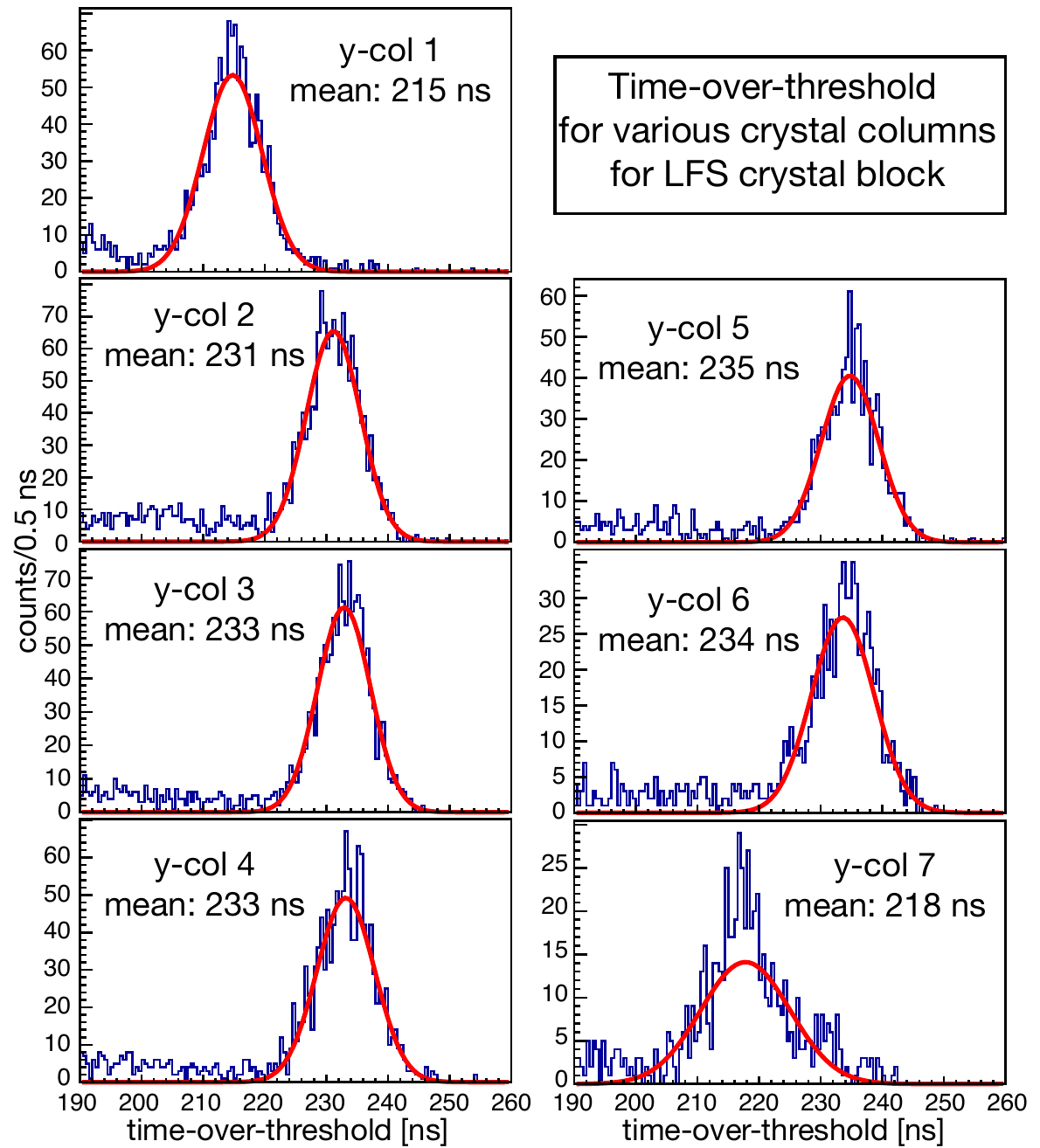}
% where an .eps filename suffix will be assumed under latex, 
% and a .pdf suffix will be assumed for pdflatex; or what has been declared
% via \DeclareGraphicsExtensions.

\caption{Typical time-over-threshold distribution for the crystal block array. As shown in fig.\,\ref{setup}, annihilation photons will enter one side of the the crystal array and can interact in any of the y-columns; y-col\,1 is closest to the source  and y-col\,7 the furthest, thus we expect to have fewer gamma interactions in the y-col\,7 just due to the radiation length of the LFS. However the point to notice is that there is less light detected in the first and last crystal.}
\label{charge-slabs}
\end{figure}

Typically MPPCs have dark counts caused by single SPADs firing, thus triggering and reading out of the Wavecatcher module should be suppressed unless there is energy deposited in the reference counter and the crystal block under test.  Signals that correspond to one or two SPADS firing have a TOT between 15\,ns and 30\,ns; these signals need to be filtered out. Thus a trigger was formed by demanding a time-over-threshold greater than 150\,ns from the reference crystal, a time-over-threshold (TOT) from one end of one of the strip MPPC to be greater than 100\,ns, and that the leading edge of the two signals were within 5\,ns. These values of TOT were chosen to be well away from the 511\,keV photopeak. In fig.\,\ref{charge-ref} we show the TOT signal for the reference counter for all triggered events; the 511 keV photo-peak is clearly visible. It should be noted that although the 511\,KeV photopeak is clearly visible, the energy scale is non-linear.  Since the Compoton edge at 340\,KeV is visible, we can derive that the energy scale that a change of 20 in TOT corresponds to 170 KeV.

The 511\,keV annihilation photon can interact in any of the seven y-columns; the time difference between each end of the strip MPPC is used to select the crystal in the y direction.  The time differences for the strip MPPCs are shown in fig.\,\ref{time-diff-1}.  We show the time difference for a single MPPC strip (upper plot), for the four strip MPPCs on one side (middle plot)  and also the average of all eight strip MPPCs (lower plot); there are clear peaks associated with each of the seven y-columns.

\begin{figure}[!h]
\centering
\includegraphics[width=9cm]{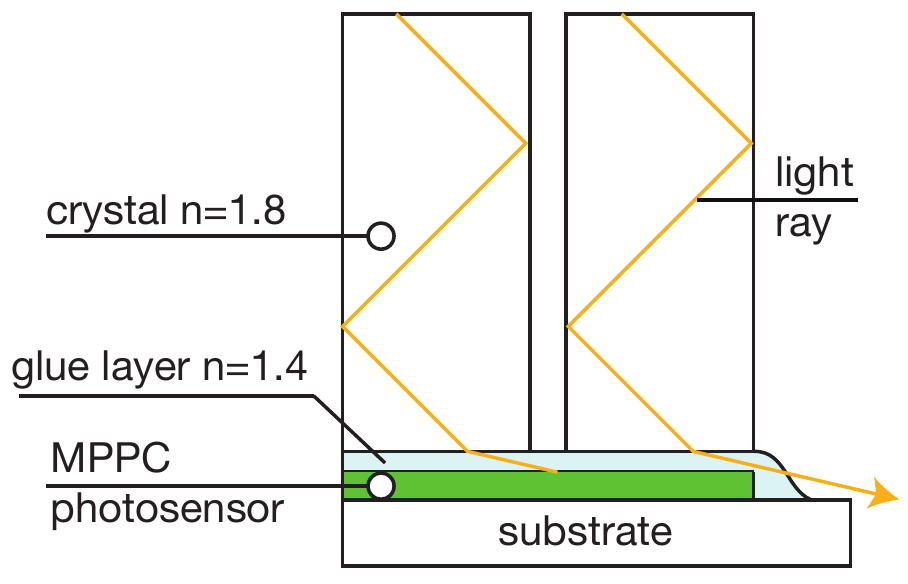}
% where an .eps filename suffix will be assumed under latex, 
% and a .pdf suffix will be assumed for pdflatex; or what has been declared
% via \DeclareGraphicsExtensions.

\caption{Schematic illustration to highlight that the glue layer (refractive index 1.4) will bend the light away from the MPPC photosensor. If the crystal is located at the edge of the MPPC - then this light will be lost.}
\label{edge-effect}
\end{figure}

\begin{figure}[!h]
\centering
\includegraphics[width=14cm]{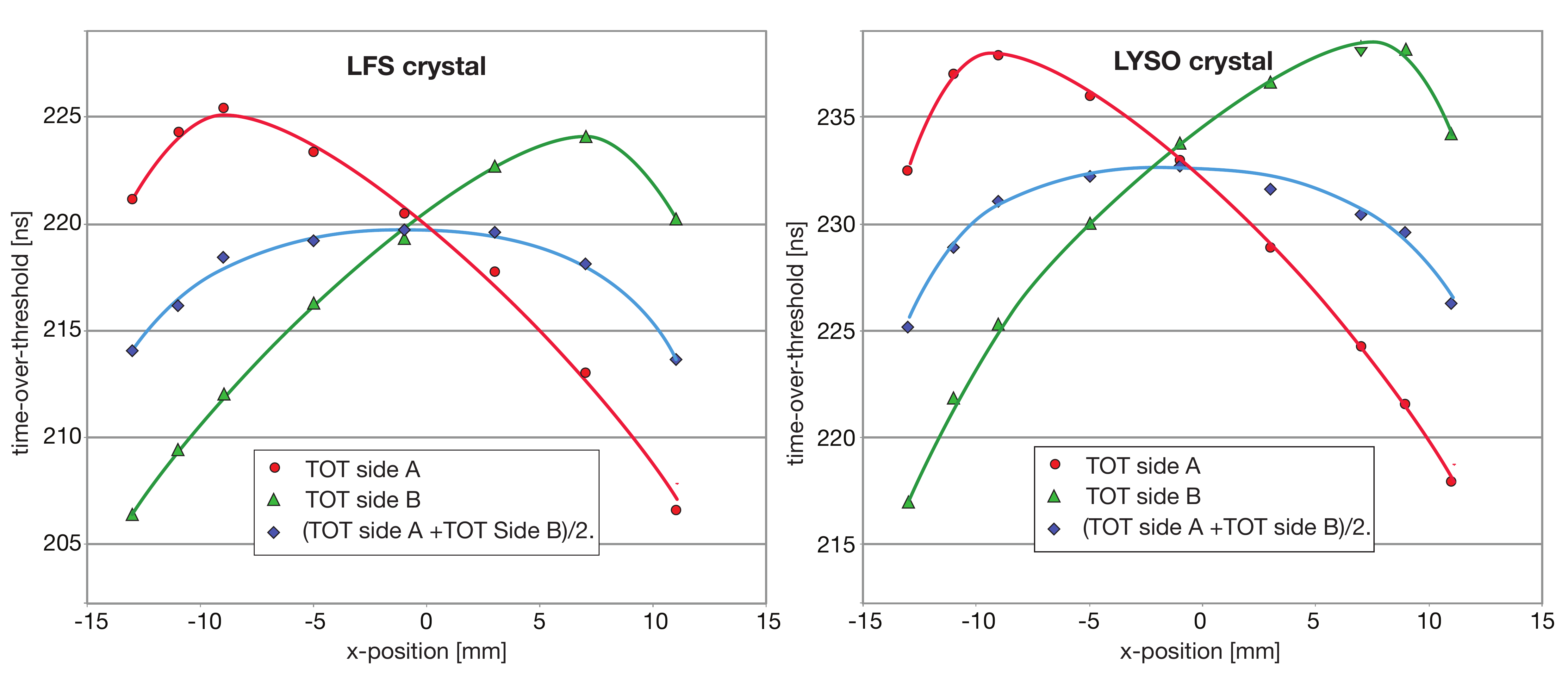}
% where an .eps filename suffix will be assumed under latex, 
% and a .pdf suffix will be assumed for pdflatex; or what has been declared
% via \DeclareGraphicsExtensions.
\caption{The time-over-threshold of the observed 511\,keV photopeak measured at each side of the crystal (side 1 and side 2) and the average of the two for a scan along the x-length of the LFS and LYSO crystal. The error bars are contained within the symbols.  The lines are to guide the eye}
\label{totvspos}
\end{figure}

\begin{figure}[!h]
\centering
\includegraphics[width=8cm]{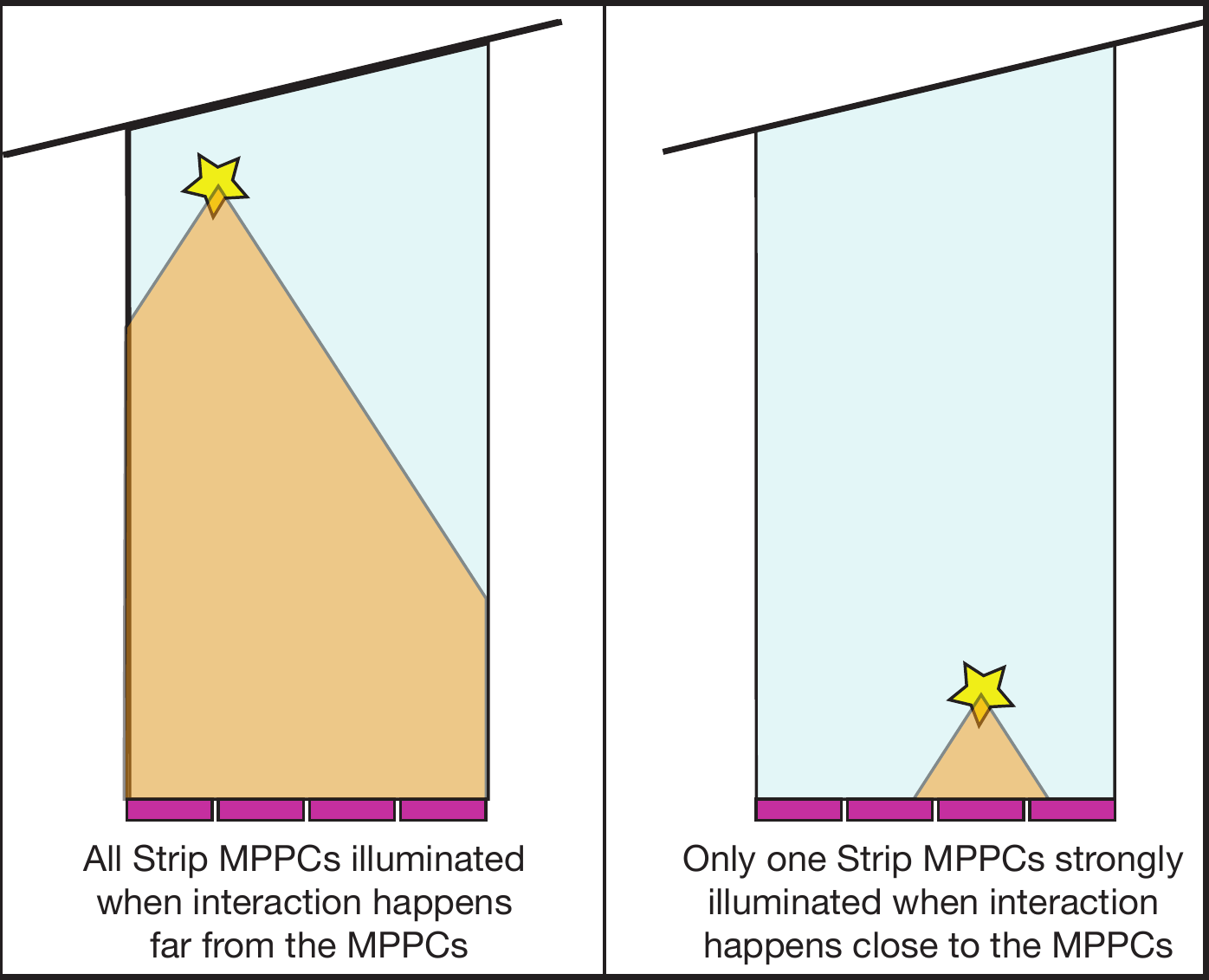}
% where an .eps filename suffix will be assumed under latex, 
% and a .pdf suffix will be assumed for pdflatex; or what has been declared
% via \DeclareGraphicsExtensions.
\caption{When the photoelectric effect interaction happens at the end of the scintillator close to the Strip MPPCs, only one strip MPPC is illuminated. }
\label{end_crystal}
\end{figure}

\begin{figure}[!t]
\centering
\includegraphics[width=12cm]{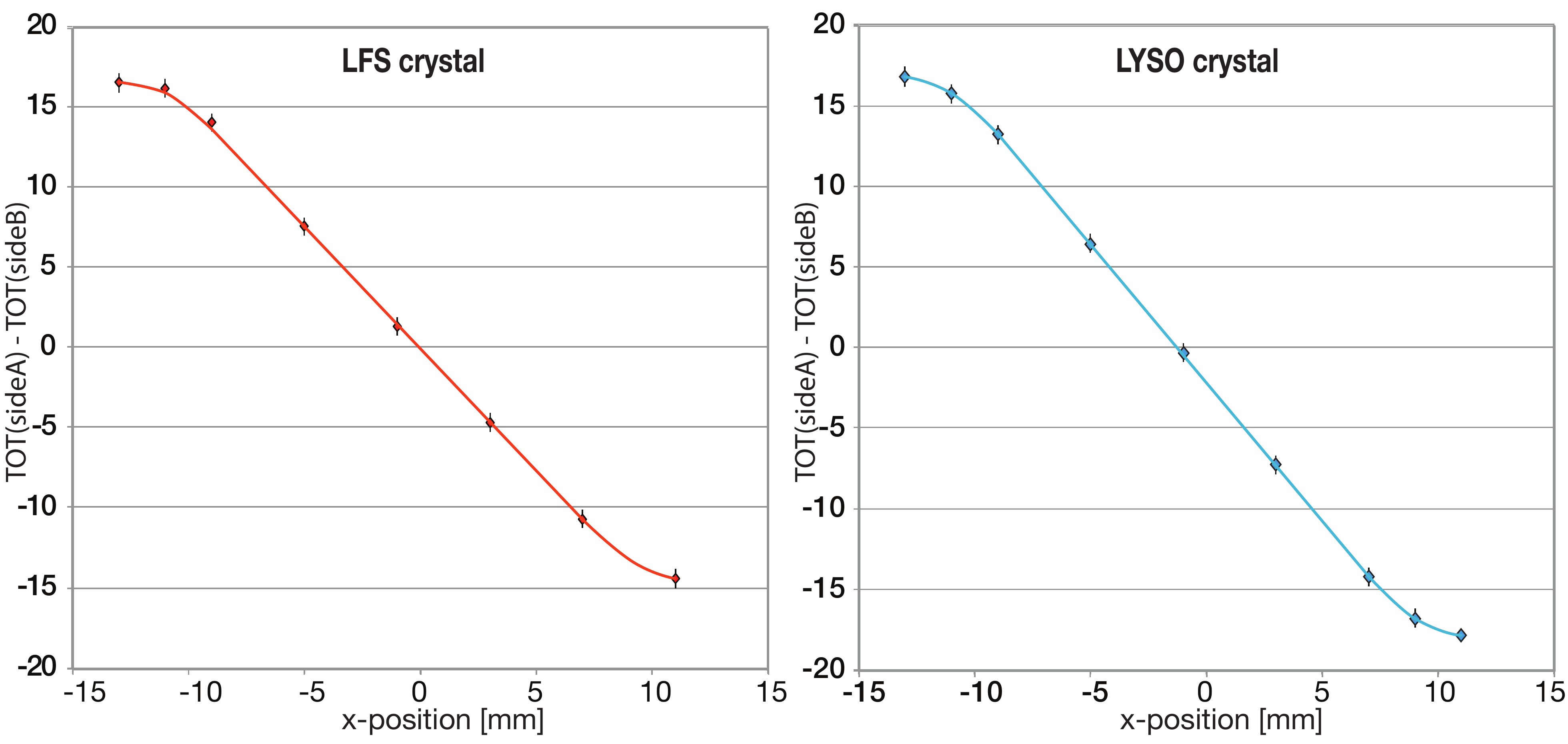}
% where an .eps filename suffix will be assumed under latex, 
% and a .pdf suffix will be assumed for pdflatex; or what has been declared
% via \DeclareGraphicsExtensions.
\caption{Difference of time-over-threshold measured at each side of the crystal block for a scan along the x-length of the crystal. The line is to guide the eye.}
\label{tot-diff}
\end{figure}

In fig.\,\ref{charge-slabs} we show the average TOT signal from the 16 channels (eight strip MPPC read out at each end) associated with the z-layer under investigation; each of the seven y-columns of the crystal is histogrammed separately; y-col\,1 is closest to the $^{22}$Na source and y-col\,7 is the furthest; thus we expect fewer events to be in y-col\,7 due to the radiation length of the LFS (12\,mm).  However it is clearly observable that the TOT of y-col\,1 and y-col\,7 are much reduced.  We believe that this is due to the refractive index of the glue (1.4) being lower than the refractive index of the crystal (1.8), thus some light is bent away from the strip MPPCs at the two external edges; this is illustrated in fig.\,\ref{edge-effect}.  During the analysis the TOT value was corrected depending on the y-column where the interaction occurred. This correction was calculated with the following procedure: the TOT peak position for the photopeak was measured for all seven y-columns (TOT-peak for y-col(i)), then for every event during the analysis, if the interaction happened in y-col(i), the TOT value was modified by the factor (TOT-peak for y-col(i))/(average TOT peak).

  \begin{figure}[!b]
\centering
\includegraphics[width=9cm]{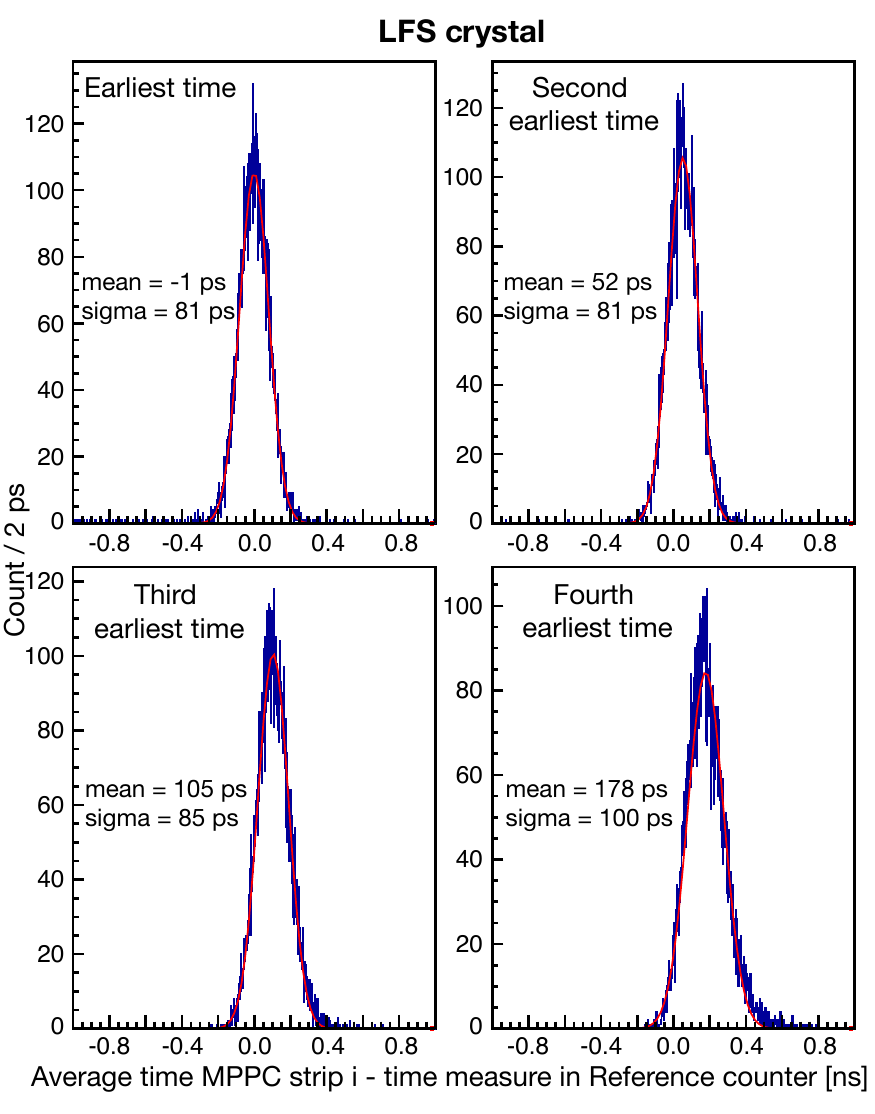}
% where an .eps filename suffix will be assumed under latex, 
% and a .pdf suffix will be assumed for pdflatex; or what has been declared
% via \DeclareGraphicsExtensions.
\caption{The time-meassurements from the four strip MPPC are sorted into time order. This data corresponds to a position close to the centre of the scintillator. }
\label{early-times}
\end{figure}

In fig.\,\ref{totvspos} we show the variation of the time-over-threshold signal (for side A, side B and the average of both ends) versus position for a scan along the length of the crystal for a particular z-layer.  Similar results were obtained for other z layers.  It is interesting to note that the average TOT decreases when the interaction happens close to the end of the crystal.  We believe that this is caused by the MPPC becoming saturated when the interaction is close to the strip MPPCs.  This is shown schematically in fig.\,\ref{end_crystal}.

In fig.\,\ref{tot-diff} we show the difference of the average time-over-threshold measured at each side of the crystal block versus position; clearly this difference can be used to determine the position of the interaction point of the 511 keV annihilation photon.  
Both sides of each crystal are read out by four strip MPPCs, with each strip MPPC read out at both ends.  The average of the time measurements from the ends of the strip MPPCs is used to determine the time of the 511\,keV annihilation photon interaction; while the time difference can be used to determine which crystal (in the y-direction) the interaction took place.  These averages are organised into a time order with the earliest time labelled as `Time 1', etc in the analysis that follows.

\subsection {Timing measurements}

The time measurements of the hits are first sorted into time order and will then be used as T1, T2, T3 and T4 (where T1 is the earliest time and T4 the latest time). However, before this time sorting, the timing of each channel has to be calibrated to account for differences in cable length and trace length on the printed circuit boards.  This is done by selecting events in the crystal that pass the TOT `photopeak' cut, and where the gamma interaction occurs in y-col\,2 using the time difference cut discussed above. Histograms of these time measurements (with respect to the time measured by the reference counter) are made; the mean of this distribution is used as the calibration offset for each strip MPPC. We show in fig.\,\ref{early-times} the time measurements after applying this offset correction and then sorting into time order for four strip MPPCs for one side of the crystal block.  

\begin{figure}[!t]
\centering
\includegraphics[width=13cm]{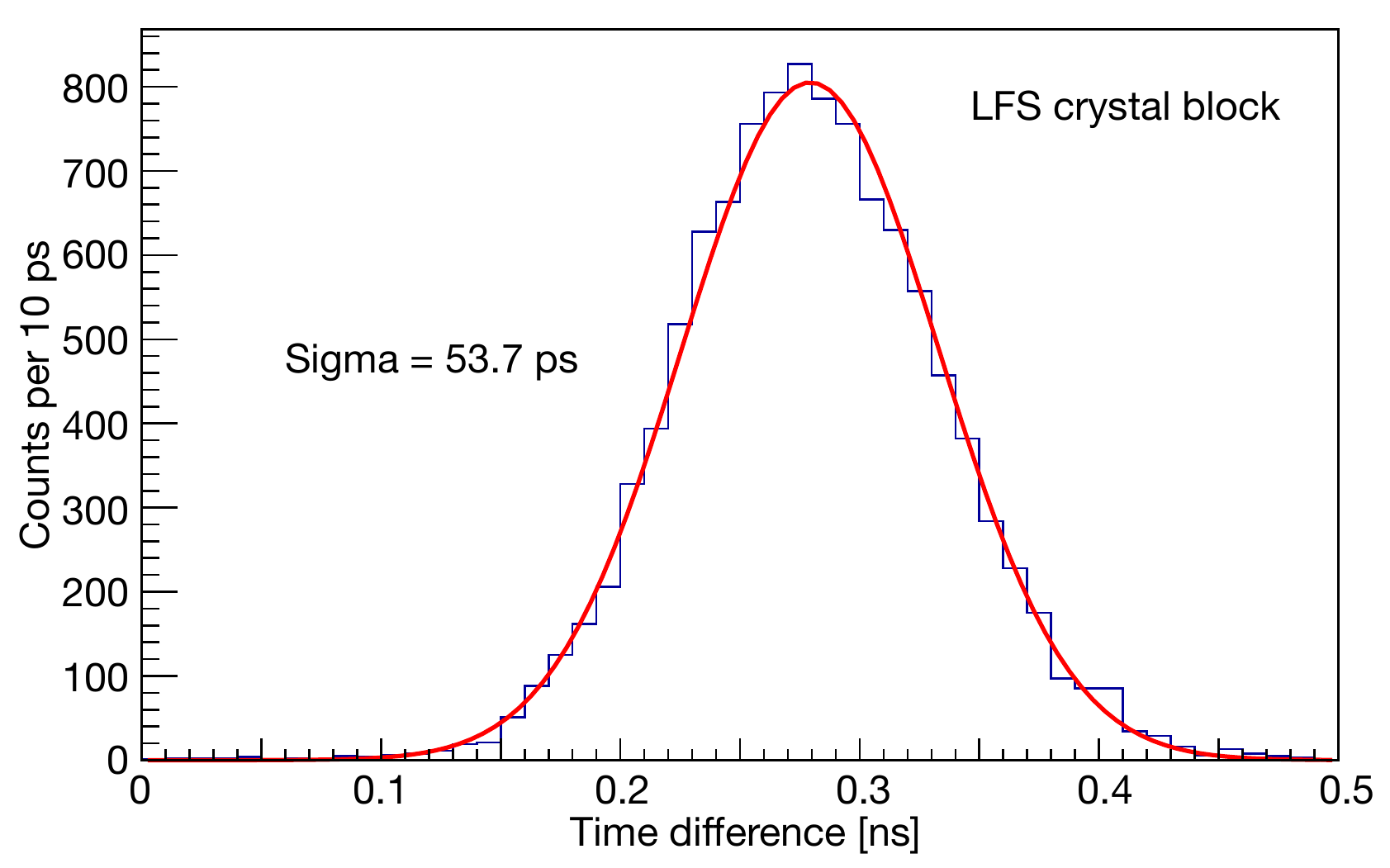}
% where an .eps filename suffix will be assumed under latex, 
% and a .pdf suffix will be assumed for pdflatex; or what has been declared
% via \DeclareGraphicsExtensions.
\caption{Typical time difference between the weighted average (see text for details) of the times measured in the strip SiPM and the time measured for the reference crystal. This is for the LFS crystal block at x-position = 4 mm.}
\label{typical-ctr}
\end{figure}

\begin{figure}[!t]
\centering
\includegraphics[width=13cm]{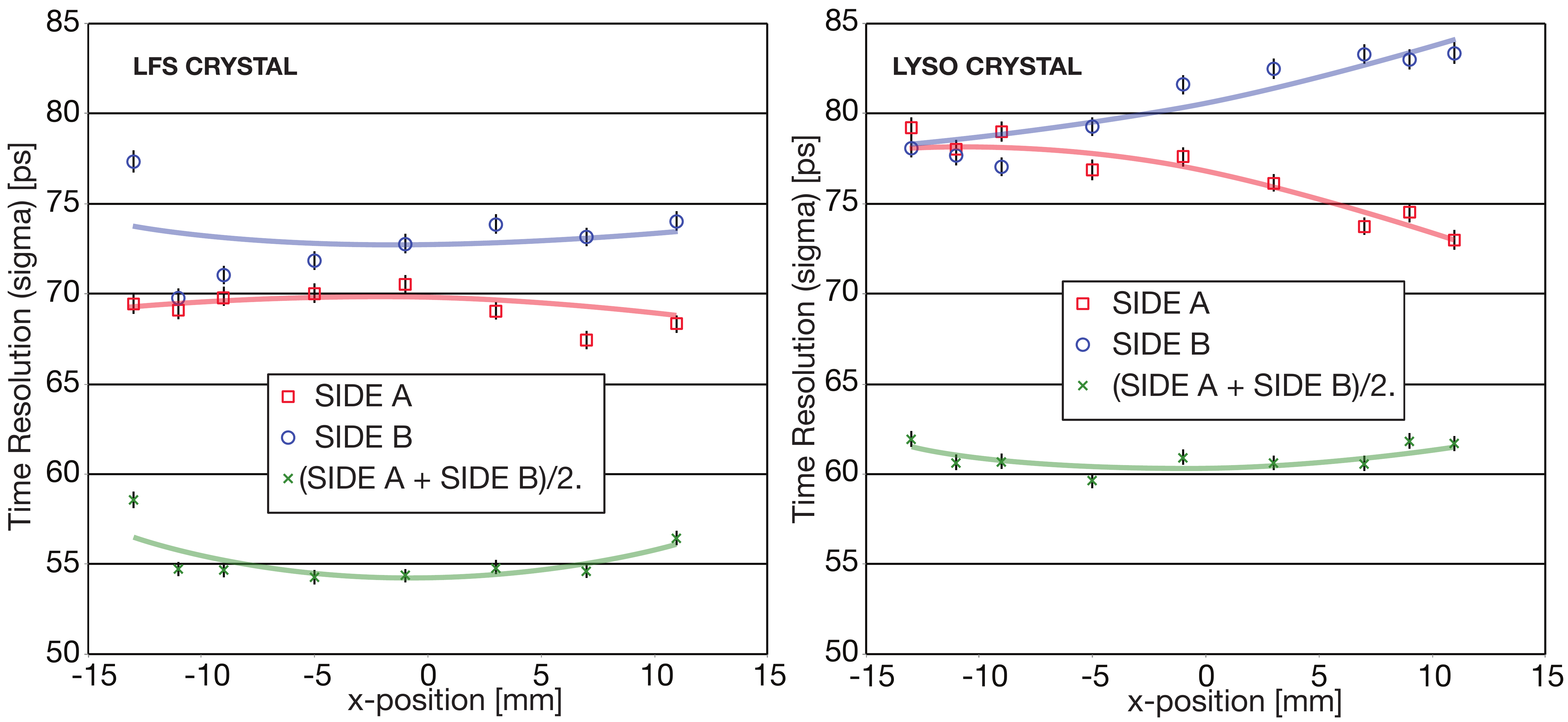}
% where an .eps filename suffix will be assumed under latex, 
% and a .pdf suffix will be assumed for pdflatex; or what has been declared
% via \DeclareGraphicsExtensions.
\caption{The time resolution measured with respect to the reference counter.  The values on each side of the crystal block array, and the event-by-event average of these two measurements for a scan along the crystal are shown for the LFS and LYSO crystals.  The lines are to guide the eye.}
\label{CTR-position}
\end{figure}

\begin{figure}
\centering
\includegraphics[width=13cm]{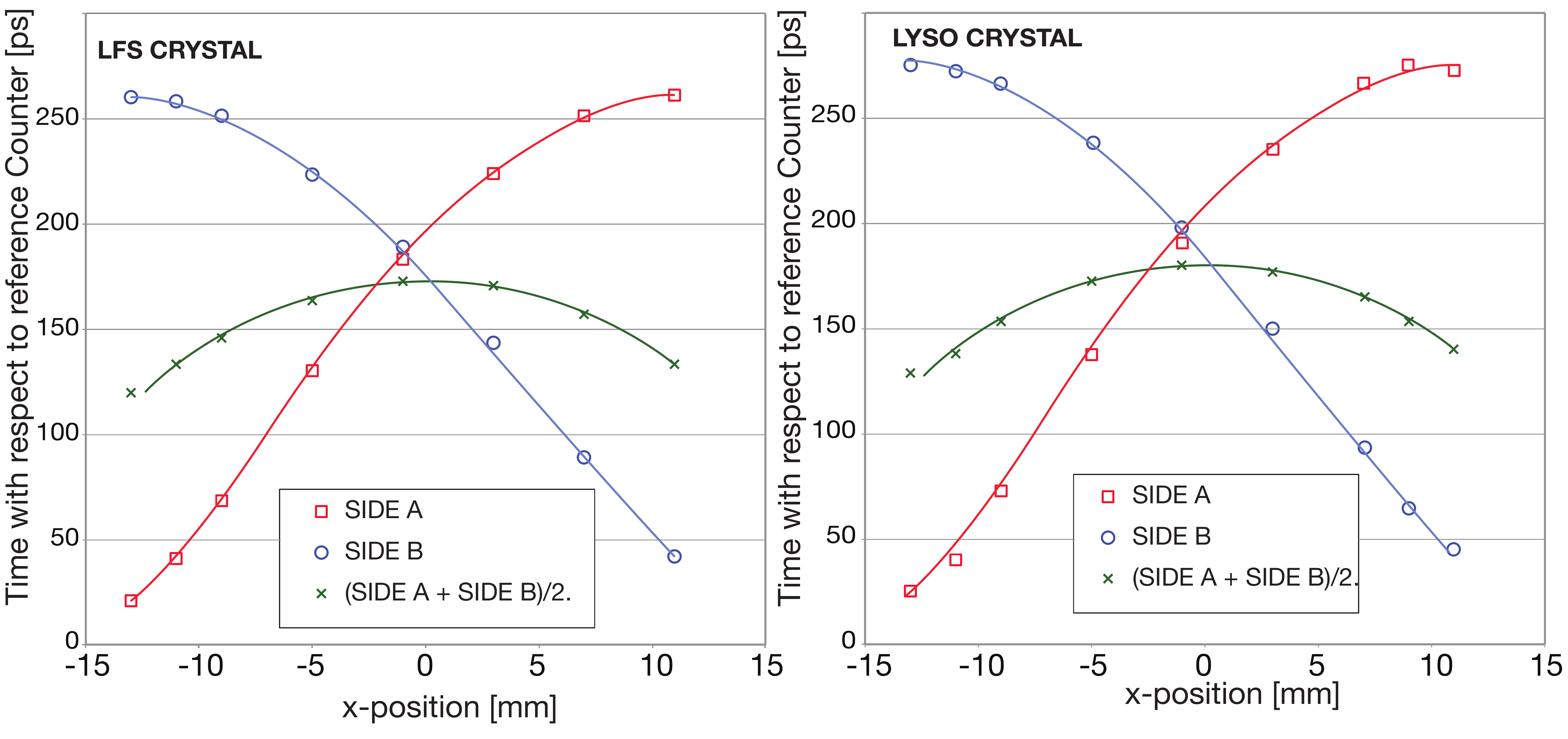}
% where an .eps filename suffix will be assumed under latex, 
% and a .pdf suffix will be assumed for pdflatex; or what has been declared
% via \DeclareGraphicsExtensions.
\caption{The time difference between each side of the crystal block and the reference counter for a scan across the crystal block for the two sides (side A and B) and the event-by-event average ((side A + side B)/2).  The error bars are contained within the symbols, The lines are to guide the eye.}
\label{time-walk}
\end{figure}

\begin{figure}[!t]
\centering
\includegraphics[width=11cm]{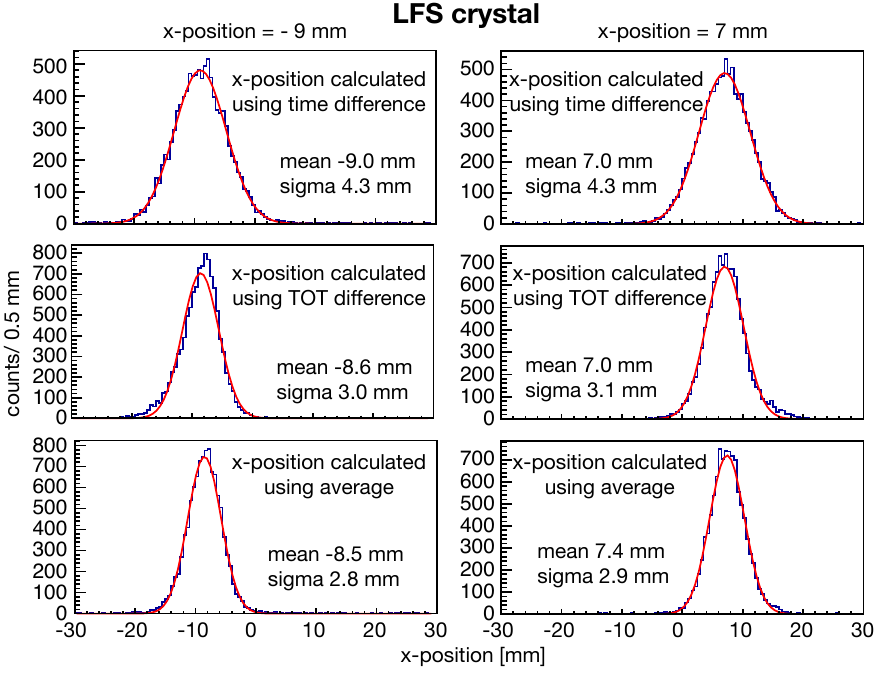}
% where an .eps filename suffix will be assumed under latex, 
% and a .pdf suffix will be assumed for pdflatex; or what has been declared
% via \DeclareGraphicsExtensions.
\caption{The calculated x-position for two different positions: -9\,mm on the left hand side and 7 mm on the right hand side. The top two plots use the time data measured on each side; the middle two use the time-over-threshold (TOT) measured on each side; the bottom two use a weighted average of the time and TOT data.  These histograms correspond to data from the LFS crystal array; the LYSO array shows similar results.  }
\label{x-coord}
\end{figure}

We investigated possible schemes of averaging these four time measurements and found, for example that using an average of the first three times gave a marginal improvement of some picoseconds with the time resolution. Further improvements were obtained by using a weighted average (T1+0.8*T2+0.5*T3)/2.3 of the first, second and third time measurement for each side and then the average of the two sides.  A typical plot of the time difference between the weighted average of the crystal block and the reference counter is shown in fig.\,\ref{typical-ctr}. We term this the time resolution; a plot of the time resolution for a scan across the crystal block is shown in fig.\,\ref{CTR-position}. 

The time resolution presented above is the time jitter of the time difference between the reference counter and the crystal block under test. We can now estimate the CTR that we would expect to be measured for two back-to-back crystal blocks.  Thus, for example, for a measured time resolution of 55\,ps (measured for the LFS crystal block), the jitter of reference counter (35\,ps ) is first subtracted in quadrature to give 42\,ps. Thus the FWHM CTR for two LFS crystal blocks would be 42\,$\times$\,$\sqrt{2}$\,$\times$\,2.35 = 140\,ps. 

Even though the LYSO has a larger time-over-threshold (as shown in fig.\,\ref{totvspos}), the time resolution is better for the LFS crystal block.  We believe that this is due to the faster decay time of LFS \cite{lfs-paper} that generates initially a higher flux of photons; the reduction of the time-over-threshold is consistent with a faster decay time.

In fig\,\ref{time-walk} we show the weighted average time measured on each side for a scan along the x-direction of the crystal; the event-by-event average between the two sides is also shown.  Using the linear part of these curves, we estimate that the inverse speed of the light pulse through the LYSO crystal is 13\,ps/mm.

The time difference between the two sides can be used to estimate the x-position of the hit; as can be seen from fig.\,\ref{time-walk}, the range in time difference is $\sim$\,540\,ps for the 25\,mm length of crystal.  Typically the sigma of the time difference for a given position is 95\,ps, thus the x-coordinate accuracy will be 4\,mm.  The x-position can also be calculated using the TOT data from each side.  Calculating the position using the time difference and the TOT difference are shown in fig.\,\ref{x-coord}.  The bottom two plots show the data using a weighted average of the time and TOT measurements (the weights are 1/sigma from the individual position plots).

 \section{Conclusions and Outlook}
 
 These measurements show that LFS crystal array gives superior time resolution than LYSO while other parameters remain the same. A LFS crystal array read out with strip MPPCs can be used to build a PET detector with a CTR of 140 ps (FWHM).  The position of the photoelectric interaction is measured with a precision (sigma) of 2.9 mm in the x direction, 2.37/$\sqrt{12}$\,=\,0.7\,mm in the y direction and 4.4/$\sqrt{12}$ = 1.3\,mm for the depth of interaction.  Clearly we need to improve the precision in the x-direction; we will focus on this aspect - but possibly better electronics will aid.
 
 There is a problem with the strip MPPC coupled to the NINO.  The characteristic impedance of the strip MPPC is 1\,-\,2\,$\Omega$ while the input impedance of the NINO is 50\,$\Omega$. We are developing a new amplifier/discriminator (the SuperNINO) that has an impedance that matches the stripline and expect better time resolution.

\acknowledgments
This work is carried out in the frame of the ERC TICAL project (grant number 338953; PI: Paul Lecoq). The authors appreciate the funding that this project receives from the EU.
%\end{linenumbers}
%\section*{References}

\end{document}